\PassOptionsToPackage{table}{xcolor}
\documentclass[manuscript,screen]{acmart}

\setcopyright{acmlicensed}
\copyrightyear{2024}
\acmYear{2024}
\acmDOI{XXXXXXX.XXXXXXX}

\acmConference[SIGCHI '25]{Special Interest Group on Computer–Human Interaction}{June 03--05,
  2018}{Woodstock, NY}
\acmISBN{978-1-4503-XXXX-X/18/06}

\graphicspath{{figs/}{figures/}{pictures/}{images/}{./}} % where to search for the images

\usepackage{tabu}                      % only used for the table example
\usepackage{booktabs}                  % only used for the table example
\usepackage{lipsum}                    % used to generate placeholder text
\usepackage{mwe}                       % used to generate placeholder figures
\usepackage{wrapfig}
\usepackage{enumitem}
\usepackage{accsupp}

\usepackage{mathptmx}                  % use matching math font

\newcommand \note[1]{{\color{black}#1}}
\newcommand{\theme}[2][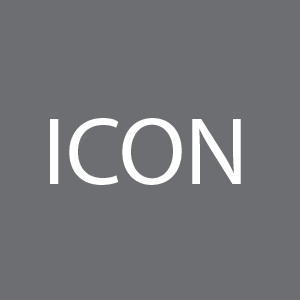]{
    \vspace{3pt}
    \noindent
    \begin{wrapfigure}[3]{l}{.075\linewidth}
    \vspace{-\intextsep}
    {\includegraphics[width=4em]{#1}}
    \end{wrapfigure} \noindent \textbf{#2}
}

\begin{document}

%% Paper title.
\title{A Tale of Two Models: Understanding Data Workers' Internal and External Representations of Complex Data}

% %% Author ORCID IDs should be specified using \authororcid like below inside
% %% of the \author command. ORCID IDs can be registered at https://orcid.org/.
% %% Include only the 16-digit dashed ID.
% \author{%
%   Connor Scully-Allison, Katy Williams, Stephanie Brink, Olga Pearce, and Katherine E. Isaacs 
% }

% \authorfooter{
%   %% insert punctuation at end of each item
%   \item
%   	Connor Scully-Allison and Katherine Isaacs are with the SCI Institue at University of Utah.
%   	E-mail: cscullyallison@sci.utah.edu
%   \item
%   	Katy Williams is with Davidson College.
%   \item 
% 	Stephanie Brink and Olga Pearce are with Lawrence Livermore National Lab.
% }

\author{Connor Scully-Allison}
\affiliation{%
  \institution{SCI Institute, University of Utah}
  \city{Salt Lake City}
  \country{USA}}
\email{cscullyallison@sci.utah.edu}

\author{Katy Williams}
\affiliation{%
  \institution{Davidson College}
  \city{Davidson}
  \country{USA}}
\email{kawilliams@davidson.edu}

\author{Stephanie Brink}
\affiliation{%
  \institution{Lawrence Livermore National Lab}
  \city{Livermore}
  \country{USA}}
\email{brink2@llnl.gov}

\author{Olga Pearce}
\affiliation{%
  \institution{Lawrence Livermore National Lab}
  \city{Livermore}
  \country{USA}}
\email{pearce8@llnl.gov}

\author{Katherine E. Isaacs}
\affiliation{%
  \institution{SCI Institute, University of Utah}
  \city{Salt Lake City}
  \country{USA}}
\email{kisaacs@sci.utah.edu}

\renewcommand{\shortauthors}{Scully-Allison et al.}

%% Abstract section.
\begin{abstract}
Data workers may have a different mental model of their data than the one reified in code. Understanding the organization of their data is necessary for analyzing data, be it through scripting, visualization, or abstract thought. More complicated organizations, such as tables with attached hierarchies, may tax people’s ability to think about and interact with data. To better understand and ultimately design for these situations, we conduct a study across a team of ten people working with the same reified data model. Through interviews and sketching, we probed their conception of the data model and developed themes through reflexive data analysis. Participants had diverse data models that differed from the reified data model, even among team members who had designed the model, resulting in parallel hazards limiting their ability to reason about the data. From these observations, we suggest potential design interventions for data analysis processes and tools.
\end{abstract}



\received{20 February 2007}
\received[revised]{12 March 2009}
\received[accepted]{5 June 2009}

%% Uncomment below to disable the manuscript note
%\renewcommand{\manuscriptnotetxt}{}

%% Copyright space is enabled by default as required by guidelines.
%% It is disabled by the 'review' option or via the following command:
%\nocopyrightspace

%%%%%%%%%%%%%%%%%%%%%%%%%%%%%%%%%%%%%%%%%%%%%%%%%%%%%%%%%%%%%%%%
%%%%%%%%%%%%%%%%%%%%%% START OF THE PAPER %%%%%%%%%%%%%%%%%%%%%%
%%%%%%%%%%%%%%%%%%%%%%%%%%%%%%%%%%%%%%%%%%%%%%%%%%%%%%%%%%%%%%%%

% \firstsection{Introduction}

\maketitle

\section{Introduction}

\note{Data workers~\cite{liu2019understanding} engaging in exploratory data analysis often directly manipulate, reference, and work with a particular data model, especially when using data analysis libraries~\cite{reback2020pandas, dplyr} and their corresponding data structures in script-based programming environments. These libraries give workers the means to view and manipulate their data but can result in errors when the data model does not conform to that user's mental model of the data. For example, a common error occurs when using a library like {\tt pandas}. Workers import their data, perform some simple operations on it, and then attempt to access a column only to be hit with a type error. The cause of the error could be an operation returning a modified structure that no longer conforms to their expectations, such as changing a primitive type, transforming a list into a scalar, or changing the object's class.

Such problems occur when there is a mismatch between that user's internal \textit{Mental Data Model} (MDM) and the external \textit{Reified Data Model} (RDM) currently represented in their code. They had expectations of the data's items, attributes, and connections between its constituent parts that did not hold when modifying and viewing the data. In the previous example, our data worker may be able to examine their code and preview the data to find the incongruities between their mental model and the reified data mode. This will fix the particular error they encountered. However, even when this debugging is successful, it takes time to solve and the mental mismatch may further inhibit efficient analysis.}

\note{Difficulties working with data models may be exacerbated as the model becomes more complex. A single table may already have a large numbers of columns containing many different types of data. Multiple tables with connections between them can increase the complexity. Beyond tables, components of the data model can be cast in other forms such as networks or geographical tiles. We refer to data composed of multiple interconnected forms as {\em heterogeneous}.}

\note{We encountered the phenomenon of people struggling with heterogeneous data models while working with a research team building their own scripting library for specialized data analysis in their domain. We were tasked with developing interactive charts to work alongside the library. When we presented them with design mock ups, they questioned why we had not chosen specific simple charts as an overview of their data. We responded that their suggestions would not represent key components of their library's data model, which they had seemingly forgotten. To us, dropping these components of their data model was particularly surprising as {\em they themselves} had designed the model for the purpose of capturing those specific components, among others. These instances demonstrate that misunderstandings of the data's form can occur even before the data is reified in code and can impact the design of data structures, APIs, and user interfaces intended to aid data exploration. Given that the data model was at the center of their library and our interfaces, we were concerned that, if the designers of the data model struggled to recall it, what did that mean for the larger audience of people who were expected to use it?}

\note{To better understand the potential struggles our target audience might have in working with a complex data model, we decided to more formally study the matches and mismatches between a reified model---the one in our collaborators' data science library---and the mental models of the people working with it. We designed and performed a qualitative study with ten data workers representing the entirety of the community working on or with the aforementioned data library at the time, save for one person. This study design allowed us to investigate differences in understanding that occurred between participants for a fixed, complex data model that all participants had spent months working with, at minimum. Seeking knowledge to help us design software that supports working with this complex data, we focused our inquiry around the following research questions:

\textbf{RQ1:} How well do people remember the aspects of a heterogeneous data structure consisting of an known form?

\textbf{RQ2:} What factors lead to remembering dimensions/aspects of the data?

\textbf{RQ3:} Given a set data structure, how diverse are the mental models of it?

}

We used multiple methods of elicitation to reveal participants' conceptions of the data, including sketching and outlining how to accomplish common analysis tasks. Using an iterative and reflexive methodology we honed and tweaked our semi-structured interview to capture emergent research questions. We then analyzed the transcripts and drawings using reflexive thematic analysis~\cite{clarke2021thematic}. 

\note{We observed significant diversity in the mental models of our participants, confirming they do not conform to the data model they had created. We identify two parallel hazards which can occur due to the complexity of mental and reified heterogeneous data models:

\begin{enumerate}
    \item If a data worker does not have an accurate representation of the data in their mind, they may be unable to do the analyses they want, regardless of the reified data model.
    \item Even if a worker knows what they want to do with data, if they do not understand the reified data model, they may be unable to express their analysis in code.   
\end{enumerate}
}

Complex, heterogeneous data presents difficulty in analysis, both on its own and in the mismatch between data workers' conceptions of it and the tools they have to work with it. We observed participants dropping dimensions or whole components of the data in their analyses and expressing difficulty placing data in the model, especially when that data was cast as metadata.

Our findings suggest data workers may benefit from additional support in learning and tracking the aspects of both data structures used in data science libraries and how those structures match to the analysis they want to perform. Furthermore, when designing such data structures and libraries, the data designers themselves may benefit from more systematic tracking their biases and assumptions about the data.

In this work, we contribute:
\begin{itemize}
    \item Empirical observations that describe domain experts' mental models of the heterogeneous data they work with, and
    \item Implications and guidance for designing user interfaces and data models to support analysis workflows using heterogeneous data.
\end{itemize}

We present the definitions and data model used in this study in \autoref{sec:background}. After discussing related work (\autoref{sec:related}), we present our study methodology (\autoref{sec:methodology}) and the themes developed from our analysis (\autoref{sec:findings}). Finally, we discuss our research questions in light of these findings (\autoref{sec:rqs}) and suggest implications for designing data models and data analysis tools (\autoref{sec:implications}).

\section{Background: Data Models}
\label{sec:background}

We define data terms used and describe the reference data model used in our
study.

\subsection{General Data Definitions Used}

    Terms for data and its representation are frequently overloaded. For
    example, a ``data model" describes some abstracted view of a dataset, but
    what does that mean practically? Is a ``data structure'' a ``data model?''
    Is a ``schema'' a ``data model?'' Does a ``data model'' encode semantics about the data? Given this ambiguity, we define how we use these
    terms:

    \textbf{Mental Data Model (MDM):} An \textit{internal} mental representation of ``data" that abstracts the specific details of said data (i.e., numbers) into a representation which is meaningful to the model holder and may not be fully consciously understood by them. The nature of that meaning is subjective and may relate to how a worker uses or interacts with the data.

    \textbf{Reified Data Model (RDM):} An \textit{external} representation of data that abstracts the specific details of the data and conveys an intentionally constructed meaning. The nature of that meaning is shared by one or more individuals and exists on a spectrum from structural to semantic. In many ways, a reified data model is similar to a schema and often exists in code as a data structure. However, a reified data model may exist outside of code solely as  a design specification or as some combination of documentation and code. The representative model in \autoref{sec:ensemble-api} is a reified data model.

    \textbf{Heterogeneous Datasets:} We define datasets as heterogeneous if
    they are abstracted into two or more interconnected basic
    data abstraction types \note{such as tables, networks, or geometry}~\cite{munzner2014visualization}. With our dataset, much
    of the data conforms to tables, however there is also a
    significant hierarchical component in it which abstracts to a tree or graph.

\subsection{{\tt EnsembleAPI} Data Model}
    \label{sec:ensemble-api}

    We interview people who are developers or users of a specific
    heterogeneous data structure, which we refer to as our {\em reference data
    model}. Our expectation was that participants would recall and use this
    reified data model in their thinking because they work directly with it regularly.
    
    The data structure is part of a domain-specific data science library
    designed to work with pandas. The domain in this case is high performance
    computing (HPC). The data represents ensembles of data collected from
    individual executions, or {\em runs}, of HPC programs. We refer to the
    library as {\tt EnsembleAPI}.

    \begin{figure}
        \centering
        \includegraphics[width=\linewidth]{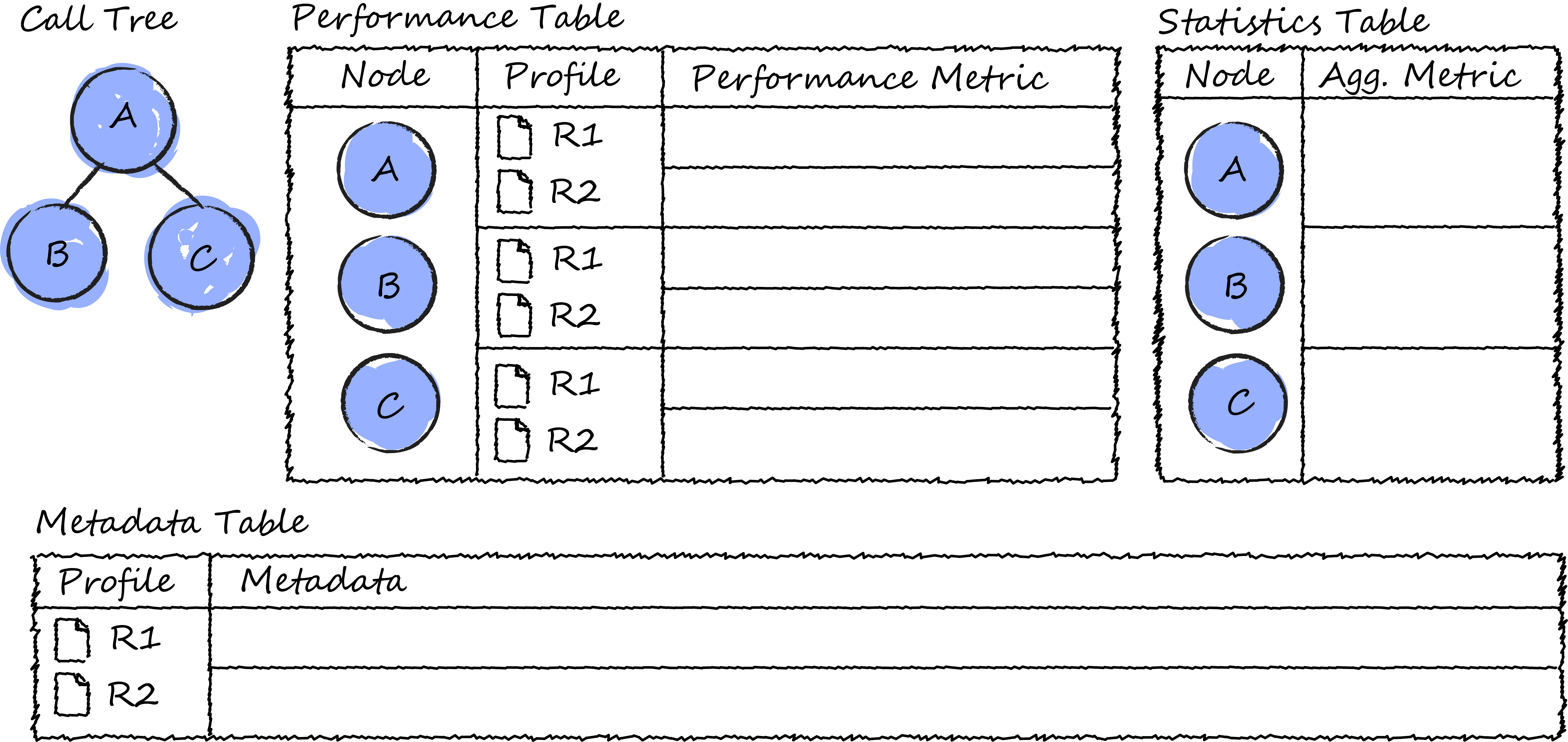}
        \caption{The reference data model of {\tt EnsembleAPI}, the project all participants were recruited from.} 
        %ALT: An image of the four components of the reference data model of EnsembleAPI: Call Tree, Performance Table, Statistics Table, and Metadata Table. The performance table has two indices, the nodes from the call tree and per-run profile ids from the metadata table. The performance table holds all measurement related data. The metadata table is per-run and holds metadata for grouping and filtering data. The statistics table has is indexed using nodes from the call tree and holds aggregated metrics from the performance table.
        \label{fig:ensemble-api}
    \end{figure}

    The reified {\tt EnsembleAPI} data model has four main components: 1) call tree,
    2) metadata table, 3) performance table, and 4) statistical table. The
    call tree is a graph data structure where nodes are functions and links
    represent caller-callee relationships. Functions can appear multiple times
    in the tree, differentiated by their path from the root (e.g., \texttt{main}).
    
    The metadata table stores data that was collected per program execution.
    It can include the context of the execution such as compiler options, input
    parameters, and hardware information. It can also include metrics
    describing the entire execution, like the total runtime. 

    The performance table stores metrics measured for each function in the
    call tree, like runtime or memory use. It is indexed on a composite key of the call tree node the data
    was collected from and the index of the run. Executions of the same program with
    different parameters will typically have rows for the same call tree node with
    different data values and possibly different metric columns. Common metrics
    include function execution time, function memory accesses, and number of data
    points processed.
  
    The statistics table stores per-function data aggregated across profiles
    and therefore is indexed by the call tree node of the function. It is
    populated when the user calls a statistical aggregation function like {\tt
    .sum()} on a particular column in the performance table.

\section{Related Works}
\label{sec:related}
We discuss related work on mental models, data abstractions, and heterogenous datasets.

\subsection{Mental Models}

Literature exploring mental models in computing can be found across many
subfields: human computer interaction~\cite{carroll2003hci,
carroll1988mental}, visualization~\cite{block2020micro, williams2023data} and
computer science education~\cite{heinonen2023synthesizing,
cetin2017reflective}. The particular "models" under scrutiny in these works
are varied and context dependant but generally seek to understand how
individuals think about a particular computing system or interface. 

Various methods to elicit mental models include interviews, diagramming tasks,
and think-aloud verbal reports~\cite{cooke1994}. Klein and Hoffman argue that
while eliciting mental models is a ``difficult, elusive, slippery task,"
exploring and evaluating best practices for describing mental models is still
a worthwhile venture~\cite{klein2008}. These works inform our methodology as we
examine mental models of data. 

Recent examinations of mental models of data have explored how users think
about their data~\cite{bigelow2014reflections, muller2019data}. The most
similar work to ours is Williams et al.~\cite{williams2023data}, which
investigated the diversity of data representations non-experts had after reading
text descriptions of small datasets without a reified data model applied. Our work
differs in that we start with a reified heterogeneous data model with which our expert
participants actively use. The model also represents a much larger quantity of
data than those used by Williams et al. Thus, our study complements that of Williams et al. by exploring a vastly different context in the space of data understanding.

\subsection{Data Abstractions and Data Models}

Data abstractions map domain-specific data to abstract data
types~\cite{munzner2014visualization, williams2023data} to communicate data design knowledge in a fashion transferable across domains.
Although not necessarily intended for use by data workers directly, understanding these
mappings can be helpful when users manipulate their base
data~\cite{bartram2021untidy} or pair visualization and
scripting~\cite{scully2024design}. 
Thus, understanding the relationship between abstractions and mental
models is critical to developing effective data exploration interfaces.

There exist typologies of singular abstractions, like tables or networks,
which are sufficient to describe many datasets. However, some data does not
easily conform to fundamental abstractions, prompting the need to design novel
ones~\cite{mckenna2014design}. Development of these new
abstractions is not trivial and is explored in the literature as
data design~\cite{bigelow2014reflections, muller2019data, walny2019data}. Even
outside of the development of novel abstractions, the assignment of data to
basic abstractions can still be tricky and may fail to
capture nuances in the data~\cite{bigelow2020guidelines,munzner2009nested}. 
Our work builds on this research by examining how data workers respond to a
heterogeneous data abstraction---one composed of multiple fundamental types. 

\note{In the HCI domain, some have argued that data models should be given more consideration in user interface design~\cite{akscyn1988data, laurel1990issues}. Others have proposed methods to use data models to simplify the process of user interface design~\cite{subramonyam2021protoai, machado2017modus, raneburger2010interactive}. In many cases, when designing an interface, the data model may already be loosely fixed by the expectations of the users or by stakeholders which provide the data. Our work expands on this by considering very explicit cases where users are expected to know a data model in some detail and would expect a hypothetical data exploration interface to conform to that model. Furthermore, our work fills a gap in the literature by tying often technical discussions of data design to the human element which such a design often impacts through data exploration interfaces.}
 
\subsection{Heterogeneous Datasets Across Domains}
\label{sec:complex-datasets}

The participants in our study were familiar with heterogeneous datasets from
one domain: performance analysis. However, other domains also have
heterogeneous datasets. Many of these datasets are associated with ``big data"
and are often more focused on the problems associated with the size of the
data and the technical problems associated with processing and accessing them in a timely manner ~\cite{anagnostopoulos2016handling, jirkovsky2014semantic,
freitas2011querying, kumar2021data}. 

Outside of these "big data"-specific manifestations of heterogeneous datasets,
scientific simulation ensembles can also be viewed as heterogeneous datasets.
These scientific ensembles can be described with multiple abstractions as they contain
"multi-variate, multi-dimensional, spatio-temporal"
features~\cite{wang2018visualization}. They are also highly complex with
numerous semantic connections binding the data together. With the continued
advance of hardware~\cite{chamberlain2020architecturally} and software~\cite{anderson2020multiphysics} developed to ensure the timely execution of
scientific simulations, scientific ensembles are increasingly common and
a focus of visualization research~\cite{wang2018visualization,
thompson2011analysis, hibbard2002visualization}. We contribute to this line of
work by investigating the mental models of heterogeneous and ensemble data and the
challenges associated with attempts to encapsulate them in reified data models and communicate them through data analysis software and user interfaces.

\section{Study Methodology}
\label{sec:methodology}

    Drawing guidance from related work, we designed and executed a qualitative
    study with developers and users of a particular reified data model to investigate how
    people conceptualize complicated data structures.

\subsection{Study Participants}

    We invited every person involved with the {\tt EnsembleAPI} project to
    participate in the study. All but one, a third technical leader,
    volunteered, resulting in ten participants total. Each were
    working on the project for at least one year. This group presents a unique opportunity to explore a breadth of interplay between mental models around a fix heterogeneous reference model.
    
    The participants of this study are all active in performance analysis for
    high-performance computing (HPC) systems. They ranged in experience and
    education from master's students to post-Ph.D. scientists. Some are more
    analysis-focused while others are more tool-focused. \autoref{tab:roles}
    summarizes the education and project role of the participants.

    .

    \rowcolors{2}{gray!25}{white}
    \begin{table}[h]
        \centering
        \begin{tabular}{l l p{4.05in}}
             & Background & Role \\ 
	     \hline
	     P1 & Student, {\it early} & develops data readers and performance data storage\\
	     P2 & Student, {\it early} & does performance analysis, develops
	     statistics-related functionality\\
	     P3 & Student, {\it early} & develops metadata-related functionality and documentation\\
             P4 & PhD & provides technical assistance to and mentors P2, P3, P5, and P7\\
	     P5 & Student, {\it early} & develops analysis workflows and integrates visualization into {\tt EnsembleAPI} \\
             P6 & PhD & technical lead of { \tt EnsembleAPI}, data model
	     designer, primary liaison to users\\
	     P7 & Student, {\it late} & provides technical assistance to P2, P3, and P5\\
	     P8 & Student, {\it late} & user of { \tt EnsembleAPI}, provides team with feedback and use cases\\
             P9 & PhD & technical lead of data collection tools that interface with the project.\\
             P10 & PhD & technical lead of { \tt EnsembleAPI}, data
	     model designer, manages developers 
        \end{tabular}
        \caption{A listing of all participants and their roles in the {\tt
	EnsembleAPI} project. \textit{Early} indicates first 2 years of
	graduate-level study.}
        \label{tab:roles}
    \end{table}

\subsection{Study Design}

    We designed a semi-structured interview to elicit participants'
    understanding of their data using multiple methods: asking them to
    describe their data, compare the importance of their data, draw their
    data, and explain how to do analysis tasks with their data. We chose these
    approaches to encourage them to think deeply about their data from many
    different perspectives. 

    Our interview document includes conditional questions for probing in cases
    where participants did not mention parts of the reference data model
    themselves. 
    We also asked a control question at the end to verify that they
    could abstract and describe a small call tree and its associated data, to
    separate ``forgetting'' the hierarchy versus not understanding it.
    
    Here are the specific questions we asked:
    \begin{enumerate}
	\itemsep=-.25ex
        \item What project do you/did you work on that uses ensembles of data?
        \item Please describe your data ensemble.
        \item Please list the main parts of the previously described data in order of importance to you.
        \item Using a pen and paper, would you sketch out the entirety of the data stored by this dataset, using any abstractions or metaphors do describe this data.
        \item Task Questions
        \begin{enumerate}
	    \itemsep=0mm
            \item Imagine we are doing a strong scaling study of how MPI threads relate to the runtime of a single function, what portions of the data would I use?
            \item If I want to compare the runtime differences between one codebase run on different architectures, can you show me which portions would be most important for that?
            \item Suppose you have a function with low exclusive time but high inclusive time. How would you find out where the time is spent? (Initially: If I want to find out which function called another function, which portion of the data is most important for that?) 
        \end{enumerate} 
        \item Contextual Questions:
        \begin{enumerate}
	    \itemsep=0mm
            \item What about the metadata? 
            \item Is there any aspect of the data that you feel is not captured by the {\tt EnsembleAPI} object? 
        \end{enumerate} 
        \item Control Question: Imagine a profile of a simple program with three functions A, B and C. Function A took 30 seconds, Function B took 5 seconds, and Function C took 100 seconds. A calls B and C. Can you draw this dataset for me?
    \end{enumerate}

    We piloted the original interview with two people who were not part of the
    {\tt EnsembleAPI} project as we knew participants were limited by the
    overall size of the project. As such, we used these pilots mostly to
    refine question wording. We thus found it useful to refine questions
    during the interview process, based on participant feedback, in particular
    Q5b and the addition of the control question after the first interview.

\subsection{Study Procedure}
    
    We began interviews with a briefing and consent affirmation. They were
    previously sent documents regarding the study procedure. We also obtained
    verbal consent to record. The interview was broken into five sections:
    initial questions (Q1-3), drawing (Q4), task questions (Q5),
    contextual questions (Q6), and the control question (Q7). The interview
    concluded with a debriefing and solicitation of any questions the
    participant had. The whole process took 20-40 minutes. 

    We conducted the interviews with a mixture of in-person and video
    conferencing. For in-person interviews, we provided paper and a choice of
    pens or crayons. Remote interviews were asked to bring their own
    implements or use a tablet.

\begin{figure*}
    \centering
    \includegraphics[width=\linewidth]{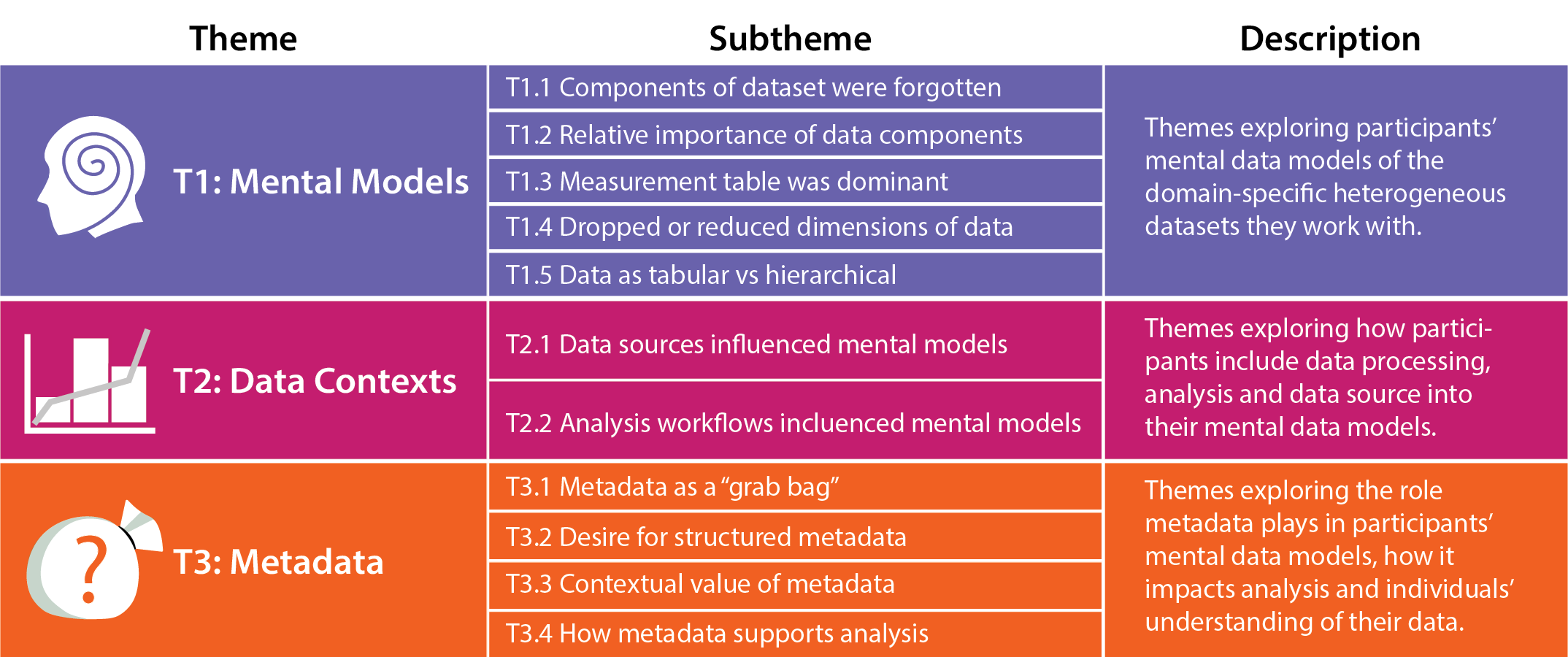}
    \caption{A summary of themes we report on in \autoref{sec:findings}.}
    % ALT: A table showing the themes we discuss in the findings section. The first column lists the themes: Mental Models, Data Contexts, and Metadata. The second column lists the subthemes associated with each theme. Mental models is broken into five subthemes, Data Contexts into two subthemes and Metadata into four. The third column has descriptions of each theme. Mental Models Description: "Themes exploring participants’ mental data models of the domain-specific heterogeneous datasets they work with." Data Contexts description: "Themes exploring how participants include data processing, analysis and data source into their mental data models." Metadata description: "Themes exploring the role metadata plays in participants’ mental data models, how it impacts analysis and individuals’ understanding of their data."
    \label{fig:themetable}
\end{figure*}

\subsection{Analysis}

    Using guidance from the psychology, visualization, and HCI communities~\cite{clarke2017thematic, clarke2021thematic, clarke2013successful, meyer2019criteria, bowman2023using} we used a {\em reflexive thematic analysis} methodology to explore and understand interview transcripts and participants' drawings. Reflexive thematic analysis describes an approach to qualitative data analysis where the ultimate goal is to develop a collection of \textit{themes} which capture patterns of meaning in a dataset~\cite{braun2006using, clarke2021thematic}. Thematic analysis is generally more constructivist or interpretivist~\cite{clarke2021thematic} in its epistemology compared to alternative qualitative analysis methodologies like grounded theory~\cite{glaser1968discovery, corbin1990grounded}.  Accordingly, thematic analysis cannot aim to describe an objective truth in full and contain it within a theoretical framework. It instead seeks to uncover a subjective reality which is a facet of the truth. Pursuant to the interpretivistic philosophy~\cite{bigelow2020guidelines, meyer2019criteria} we adopted for this research, reflexive thematic analysis enabled us to bring our own perspectives and personal knowledge about the subdomain of HPC and the participants to the analysis process. 

    We used a two phase analysis, one using \textit{deductive} coding to identify which facets of the reference data model were communicated and one using \textit{inductive} coding to develop themes beyond what could be captured by the deductive codes. In the deductive phase, we used the components of the reference models as deductive codes. Three coders participated in this phase.

    The coding in the inductive phase was performed by a single coder. We chose to use a single coder following the guidance of Clarke and Braun~\cite{clarke2021thematic}. They say that one coder in Reflexive TA, ``is normal practice and\ldots good practice.'' In Reflexive TA's interpretivist grounding one subjective viewpoint can uncover interesting aspects of a dataset with no risk of ``missing'' an objective truth. Multiple coders can provide ``a richer and more nuanced understanding of data'' but are not necessary. An additional author from the deductive phase met regularly with the single coder to discuss and help refine these codes and their development into themes. (See supplementary materials for all codes and themes developed from these rounds.)

\begin{figure}
    \centering
    \includegraphics[width=.9\columnwidth]{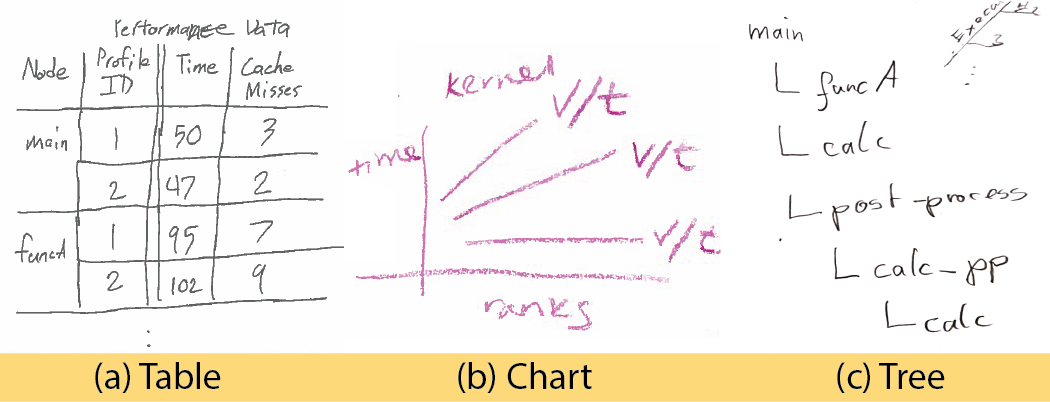}
    \caption{Common visual idioms use by participants to represent the data
    model components. Tables were drawn with and without lines. Trees were mostly indented, but some were node-link. Line charts were often used to express the entirety of the dataset. (See supplementary materials for all drawings.)}
    %ALT: Figure showing three excerpts of participant's sketches. On the left, a hand drawn table labeled above "Performance Data". This table has 4 columns. From left to right they are labeled, "Node," "Profile ID" "Time" and "Cache Misses." A vertical double line separates the left two columns (indices) from the right two (measurements). Synthetic data populates the cells below them. A label below says "a. Table." The middle sketch shows a hand drawn line chart with multiple lines. The chart is labeled above with the word "Kernel." The left axis is labeled "time," the bottom axis is labeled "ranks." The multiple lines in the chart are labeled "v/t."  This sketch is labeled "b. chart" below. The final sketch is of a hand drawn indented tree. The root is labeled "main" with three children: "funcA", "calc" and "post-process". "Post-process" has one more child "calc_pp" and grandchild "calc".  This sketch is labeled below "c. Tree." 
    \label{fig:sketches}
\end{figure}

\subsection{Statement of Positionality}

  All coders have published in both data visualization and HPC research venues
  and are familiar with the data represented by {\tt EnsembleAPI} and
  performance analysis with ensemble member datasets. The first author and
  primary coder has worked with the researchers leading {\tt EnsembleAPI} for
  four years, including several months embedded at their institution. Both the
  first and last author (primary and secondary coders) were involved in the
  {\tt EnsembleAPI} project. 

  Our expertise in both domains permits us to interpret the highly technical
  and often domain-specific communication of the participants.  The first and
  last author additionally could interpret using knowledge about the project
  itself and a deep understanding of the participants' roles in it. 
  
  Though our domain expertise and connection to the {\tt EnsembleAPI} project
  make us uniquely equipped to interpret our participants' responses, this
  familiarity may also have encouraged participants to be less comprehensive
  in their descriptions. Whenever we noticed a participant attempting
  to lean on an assumed mutual familiarity, we probed further.

\section{Findings}
\label{sec:findings}

We developed three main themes, composed of eleven sub-themes. They are
summarized in \autoref{fig:themetable} and described below.

\subsection{Mental Models}

These subthemes describe our observations regarding the content of our
participants' mental data models.

    \begin{figure}
        \centering
        \includegraphics[width=\linewidth]{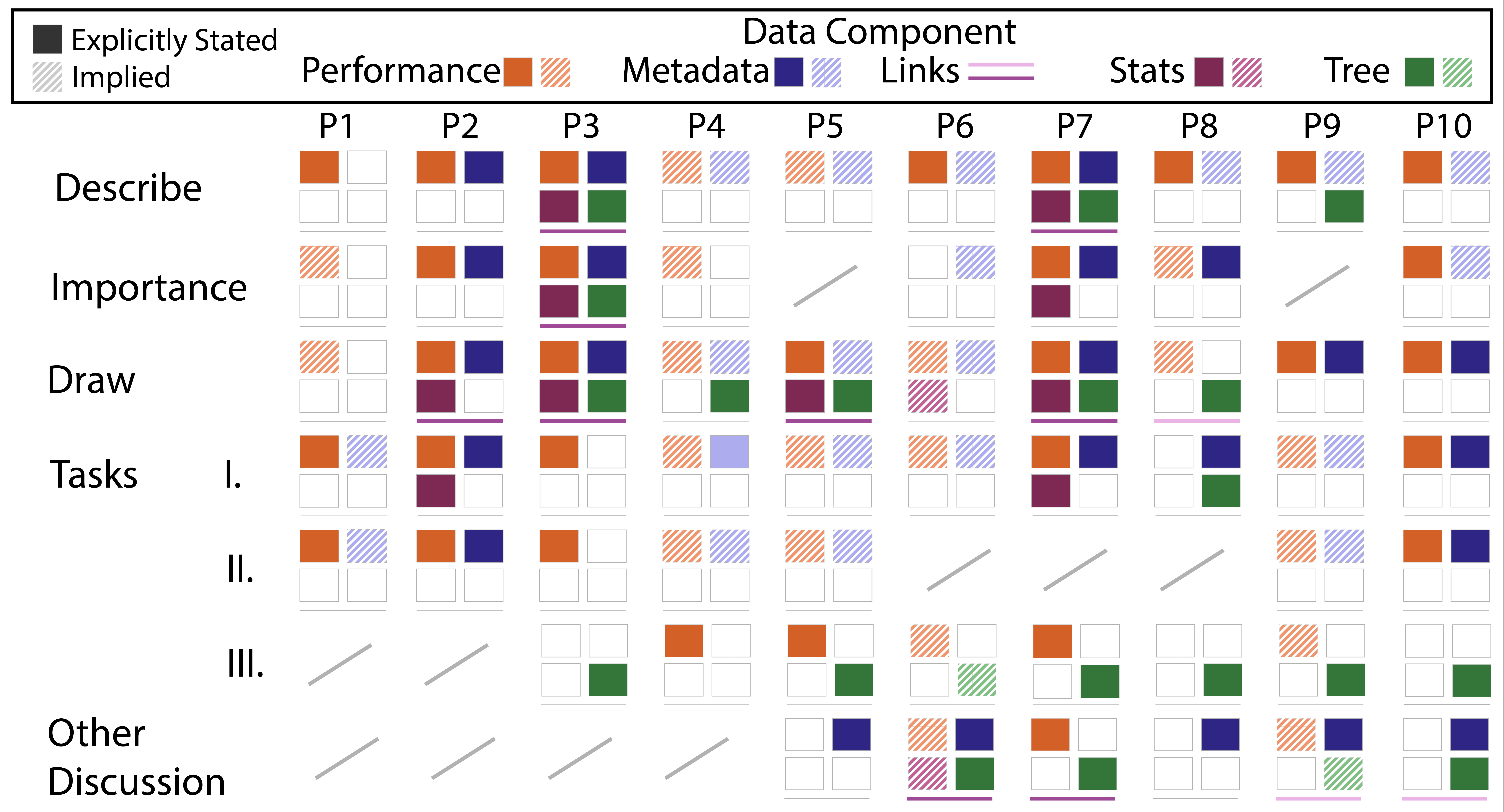}
        \caption{The "timeline" of recalled data components per participant. The left axis lists interview questions in the order they were asked, top to bottom. Initial recall was limited when we asked participants to describe their data. Additional elicitation methods (drawing, task questions) lead individuals to recall more and think about their data in different ways.}
		%ALT: A visual table showing when participants recalled certain components of the data model. Each "cell" in the table has 4 squares which are filled with a solid color if a component was explicitly mentioned and a striped color if a component was implied. The "rows" are labeled from top to bottom "Describe, Importance, Draw, Tasks I, Tasks II, Tasks III, and Other Discussion." This is the order in which questions in the interview. The columns are labeled from left to right with the Participant ids, P1-P10. Overall, recall of data parts was limited for the first question, with many participants mentioning only performance data and implying the presence of metadata. Later questions (drawing and tasks) prompted participants to mention more data components explicitly. P3 and P7 are prominent outliers however as both of them recalled all data parts immediately and continued to mention them in the next few questions.
        \label{fig:timeline}
    \end{figure}

    \theme[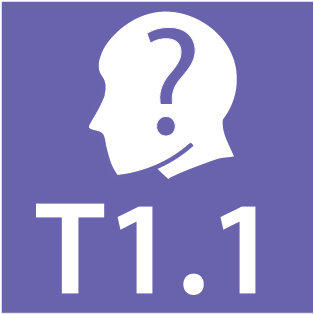]{Parts of the heterogeneous data were forgotten.}
    \label{sec:forgotten}
	There was evidence that participants regularly forgot or dropped parts of the heterogeneous dataset compared to the reference data model (\autoref{fig:ensemble-api}). The reference data model consists of a performance data table, metadata table, call tree structure, statistics table, and the relationships among them. \autoref{fig:timeline} summarizes when in the interview and how explicitly participants recalled these components. 

	Performance measurement data was the most widely recalled (see also, T1.3). Several
	participants (P1-P3, P7, P9) both explicitly described ``performance
	data'' as part of their dataset, using that specific phrase, and drew
	it (e.g. Fig.~\ref{fig:sketches}(a)). Among the remaining participants, P4,
	P5, and P8 spoke about performance data tangentially and also drew it.
	They cited common examples of performance data ("counters",
	"timestamps", "metrics") (P4, P5, P8). P6 and P10 were less inclined to
	refer to performance data in the abstract. Instead they alluded to it
	when describing analysis processes. P10 discussed and drew their data
	more as a list of metrics and associated charts. P6 implied that some
	data was ``performance data'' by placing it on the x-axis of example
	line charts.

	After performance data, recollection and mention of the other data
	parts was more varied. \emph{Metadata} was explicitly mentioned by
	five participants when asked to describe their data and subsequently
	included in their drawings (P2-P4, P7, P10). P5, P6, P8, and P9 alluded to
	metadata by implying a grouping or ordering of their data using a
	variable typically stored in metadata. The others in this group only
	included metadata as an axis on a chart with the exception of P5 who
	did not structure their drawing in a way to imply they wanted to
	reference to metadata. P1 did not recall the metadata at all despite
	our probing.  We explore participants relationships to metadata in
	more detail in \autoref{sec:metadata-theme}.

\begin{figure}
        \centering
        \includegraphics[width=\columnwidth]{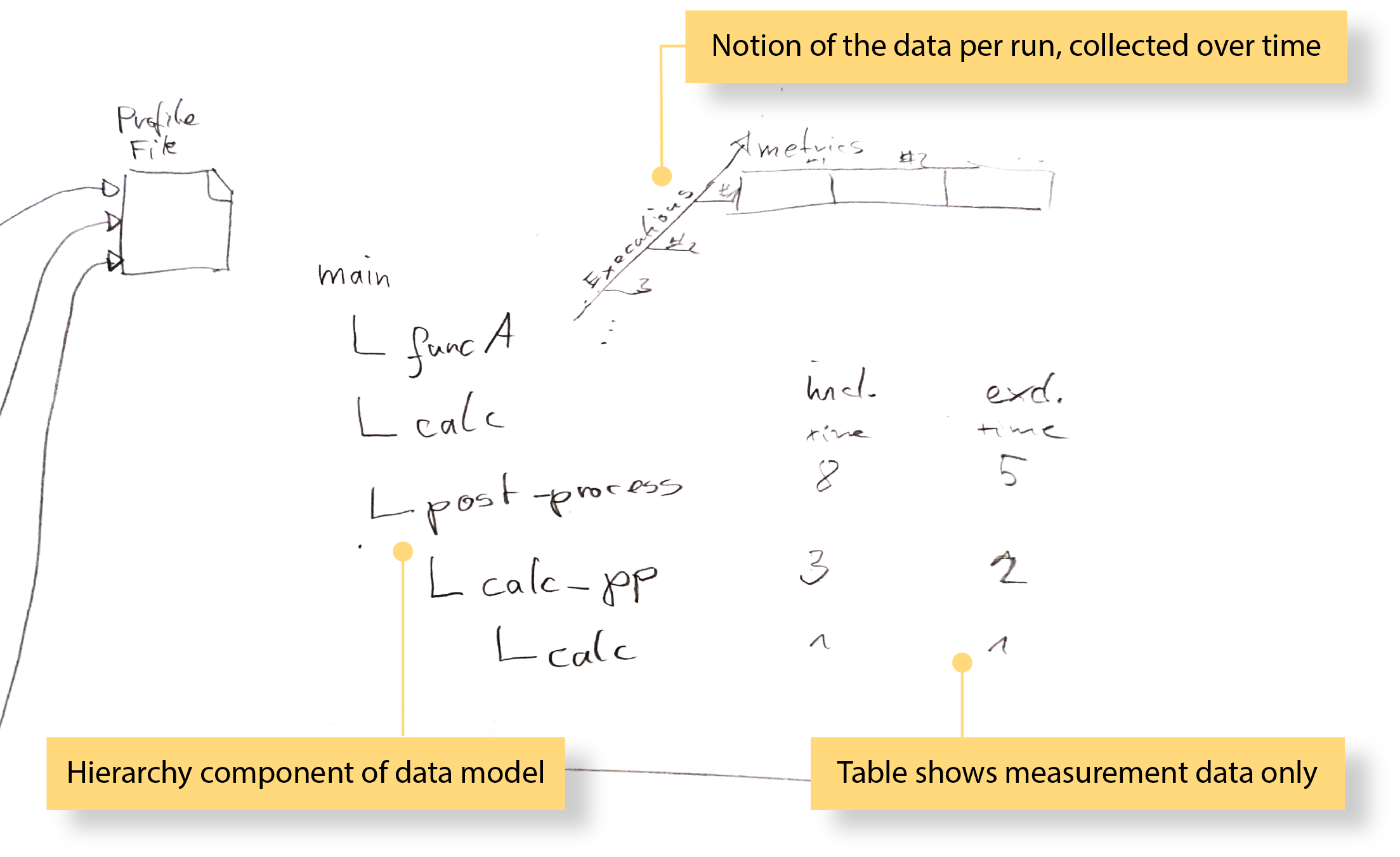}
        \caption{An excerpt of P4's sketch. This sketch exemplifies how participants rendered hierarchies. The element in the top right corner shows P4's conception of a single run in the ensemble dataset: highly granular temporal data is aggregated to produce the tree drawn beneath. This element is an example of the diversity of observed models.}
		%ALT: A sketch of portions of a heterogenous dataset from P4s perspective. In the top right, hand-drawn rows of 'metrics' are connected perpendicularly to an arrow labeled 'executions.' Each row of metrics is labeled with a number from 1-3. The paper authors annotate this element, saying 'Notion of the data per run, collected over time.' Below this element is a hand-drawn indented tree labeled with function names from a hypothetical program. To the right of this tree is a small table with two columns 'inclusive time' and 'exclusive time'. The data in the 'cells' of the table are aligned to the nodes in the tree and the data is synthetic. The tree is annotated by the paper authors 'Hierarchy component of data model.' The table is annotated by the paper authors 'Table shows measurement data only.'
        \label{fig:P4}
    \end{figure}

	Recollection of the \emph{Call Tree} was highly variable. P3, P4, P7, P8, and P10 included the call tree in both their description of the data and drawings
	(e.g. Fig.~\ref{fig:P4}). P10 did not include a tree in their
	initial drawings, but revised their drawing when reminded of the tree
	by a task-focused interview question. Participants who partially
	remembered the tree (P5, P6, and P9) referenced it either directly
	with the name "call tree/ call graph" or by evoking relationships
	between call sites, nodes, or regions. Among these individuals, P5 and
	P8 are the only ones who included a tree visualization in their
	drawing of the data. P1 and P2 did not include the call tree in either
	their data description or drawing.

\begin{figure}
        \centering
        \includegraphics[width=\columnwidth]{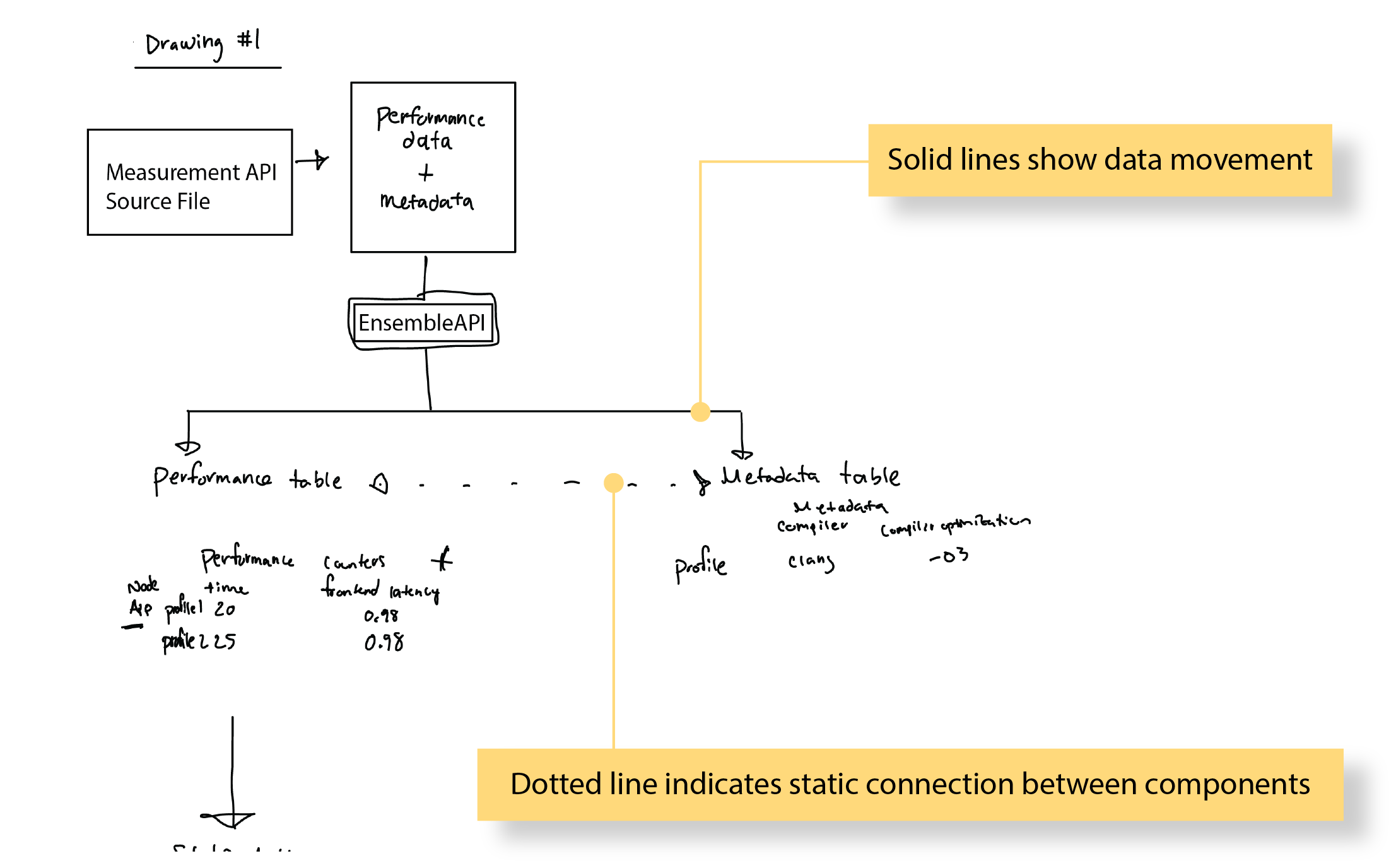}
        \caption{An excerpt of P2's sketch. P2's exemplifies usages of arrows across participants' drawings. Arrows were most commonly used to indicate a workflow of data movement between data components or from the data source. The were also used to denote links between parts of the data.}
		%ALT: An excerpt of P2s sketch of a heterogenous dataset. A solid line traces a path from a word '.cali' denoting a filetype, to a large box labeled 'performance data and metadata'. A line emerges from this box out to another labeled 'EnsembleAPI.' From this box a forked arrow emerges pointing to different components of the dataset labeled 'performance table' and 'metadata table.' Beneath each of these labels are tables with no lines separating cells of synthetic data. A dotted line connects these two components. Two annotations highlight the usage of different lines in this sketch. One pointing to the solid lines says 'Solid lines show data movement.' The other, pointing to the dotted line says 'Dotted line indicates static connection between components.'
        \label{fig:P2}
    \end{figure}

	For relationships, participants were much more inclined to verbally
	express links between components, compared to drawing them. From the
	collected sketches, we observed the use of arrows and lines in the
	drawings of five participants (P2, P3, P4, P5, and P7) (e.g. Fig.
	\ref{fig:P2}).  This usage suggests links between data parts;
	however, in most instances (P2 [forked solid-line], P4, P5 and P7),
	these lines indicate a workflow or transformation in the data and not
	static connections between data. P2 expressed connections between data
	with a dashed-line and P3 is the only participant who exclusively
	used lines to show connections between data.

	All participants except P2 used explicit database terminology when
	describing how items stored in tables relate. P3
	referenced a Database Management System saying, ``performance data
	acts as sort of the main table, if you look at it from\ldots our relational database management model."
	P5 used the term ``view" in a SQL-like fashion to describe a derived
	table. P7 discussed indexing extensively and even describes {\tt
	EnsembleAPI} as ``almost like a graph database." 

	Among the remaining participants, P1, P4, P6, P8, and P10 expressed
	relationships in more abstract terms that implied associations between
	data parts without specifying them in great detail. For example, P10
	related their categories of variants and tunings using concepts of
	membership by saying ``within each variant, you have different
	tunings." The details of how exactly a tuning is ``contained" in a
	variant was left ambiguous.
        
	The statistics table was most forgotten. However, it exists for
	derived data and is not populated on collection. Thus, we chose
	not to consider it further in our analysis.

    \theme[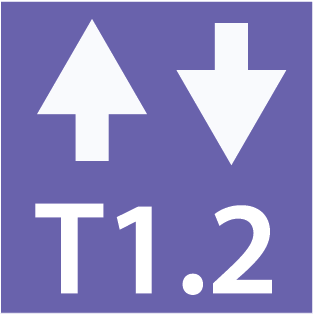]{Participants gave varying reasons for what made
    a component of the data "most important."}
    \label{sec:importance-reasons}
	Eight participants (P1-P5, P7, P8, P10) stated which data components
	were most important along with either explicit or implicit
	justifications. P6 and P9's drawings and descriptions did not exhibit
	rankable components. The most common justifications for importance were how the component
	impacts insight generation and their personal interaction with the
	component.

	Impact on insight generation was the most common justification (P4,
	P5, P7, P8, P10) and often connected to the performance data. P4
	suggested that importance "boils down to the question you are
	investigating." P7 said:

        \begin{quote}
        Yeah, so the most important is definitely the performance data.
	    That's your actual stuff that you have to analyze. --- P7 (Student, Technical Advisor)
        \end{quote}

	P10 indicated domain-specific metadata as most important. They said,
	"[they are] really helping us understand where we are running\ldots
	and what optimizations we might have done to that kernel."

	Personal interaction was cited by four participants (P1-P3, P8). P1
	said "I order [the components] because when I'm developing, the
	significance of what I'm interacting with most." P2 cited the
	performance data as most important "because most of the work that I do
	is on the performance side." P3 said "I've worked mainly with the
	metadata table. So that's why it took precedence."

    \theme[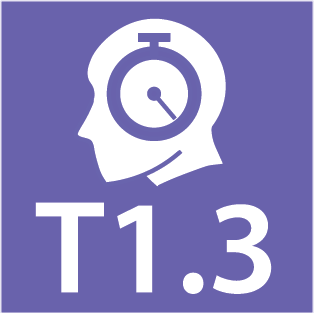]{Participants' mental models coalesced around the table most associated with measurement.}

	Seven participants (P1-P3, P6-P10) implicitly or explicitly cast the
	``performance data," as quintessential to their mental models. The
	performance data is where most of the measurements in both the
	ensemble and each of the ensemble member datasets are stored. The
	collection of this data has been a major technical area in HPC for
	decades.

	This data was characterized as ``primary" or more ``real" than the
	other parts of their dataset. Participants used modifiers like
	``actual" and ``main" when describing it.  P1 called it ``the data
	itself", P2 and P4 called it the ``actual outputs" and ``actual
	metrics", respectively. P3 called the performance data table ``the
	main data table" and discussed other data parts in relation to it. 

	Participants who drew charts(e.g. Fig.~\ref{fig:sketches}(b)) demonstrated 
	an implied interest in the measurement data, creating a visualization
	which emphasized the salient characteristics of that data.
	Metadata was used to organize the charts as the independent variable
	to the performance data's dependent variable.

    \begin{figure}
        \centering
        \includegraphics[width=\columnwidth]{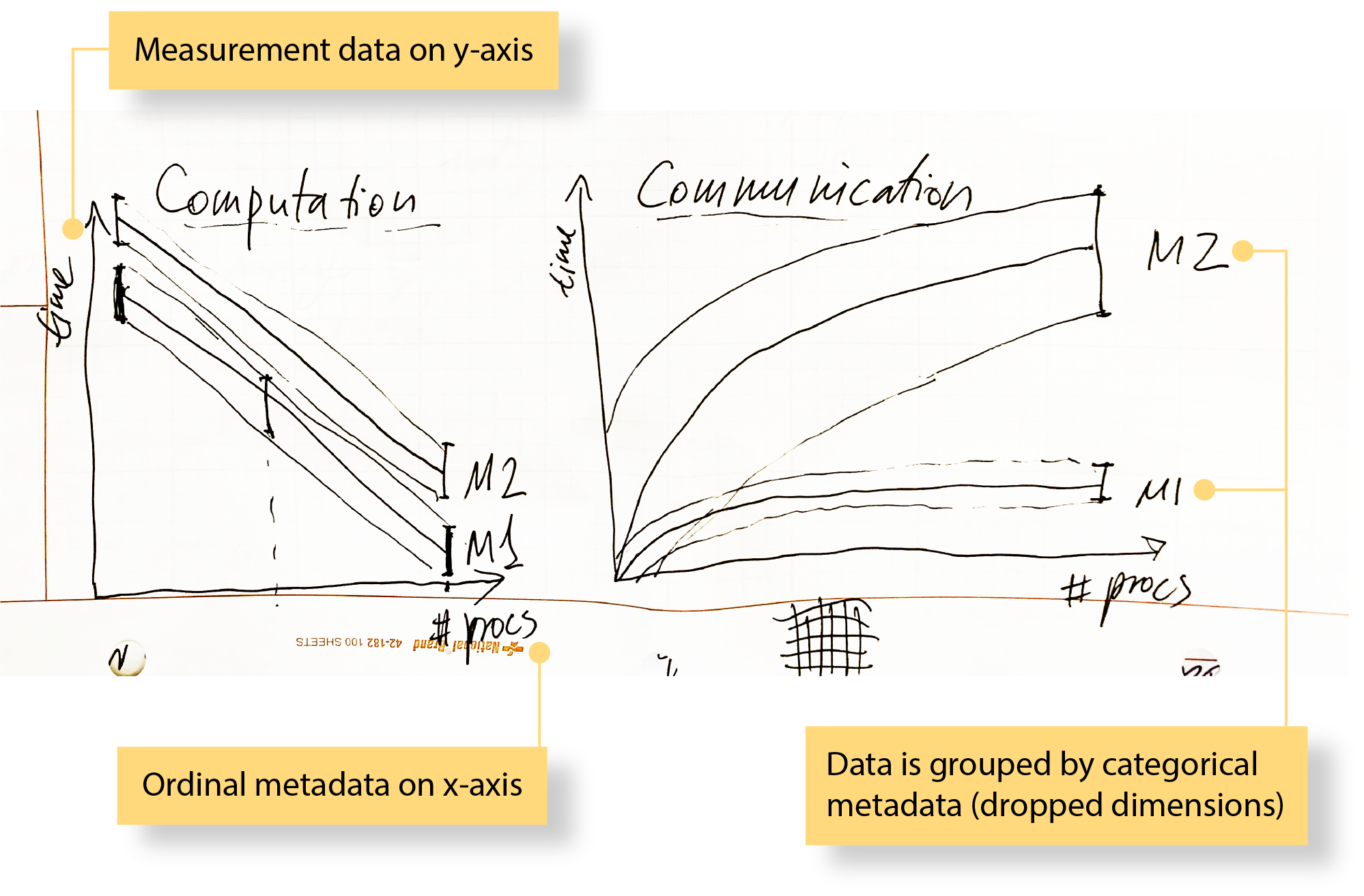}
        \caption{P6's sketch exemplifies the mental models held by analysis focused participants. These charts imply the presence of data components by showing variables from different components on each axis. P6 implicitly aggregated the data by drawing two lines for comparison, this demonstrates how dimensions are dropped or reduced as participants think about their data. }
        \label{fig:P6}
		%ALT: An hand drawn image of two line charts side by side showing synthetic data to explain how P6 thinks about their dataset. The left chart is labeled 'Computation,' the right, 'Communication.' Both charts have the same axis labels. 'Time' for the y-axis, '#procs' for the x-axis. The two lines on the left chart start in the upper left and descend to the lower right. Parallel lines running adjacent to each line denote uncertainty and error margins. The lines are labeled M1 and M2. The right chart's lines curve up and to the right depicting clear logarithmic curves. They are labeled M1 and M2 and have the same 'error bars' to show uncertainty. Three annotations on the sketch (added by the paper authors) highlight interesting elements. The first, says 'Measurement data on y-axis.' The second says 'Ordinal metadata on x-axis.' The third points to the M1 and M2 labels and says 'Data is grouped by categorical metadata (dropped dimensions)'.
    \end{figure}

    \theme[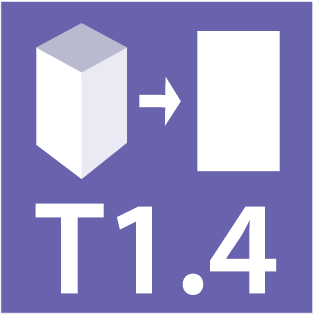]{Participants dropped or reduced dimensions of the data.}
	Throughout our interviews, and in informal discussions prior to the
	interviews themselves, some participants---mostly those more
	senior---were prone to outright drop or implicitly aggregate data
	dimensions when describing their data. 

	While often these dropped dimensions were intentional, twice in our interviews where this dropping happened
	accidentally.

	P6, when discussing the plot they drew in their sketch, said:

        \begin{quote}
            \ldots each data point in here is basically\ldots one run. And then the error bars? Well, no, that's not a one run, it's an average of the runs. And the error bars are also the average for the different runs right? --- P6 (PhD, Technical Lead)
        \end{quote}

	P6 worked backwards from their desired analysis plot to describe the
	data components. They forgot some of the steps which were required
	to transform the data from its ``raw" representation in the reified data model
	into their desired chart. 

	P7 forgot a dimension in their response to question (task) 5a. When
	explaining how a particular metadata variable
	can be used, they said:

        \begin{quote}
            So you'd have in this case, in fact, you probably have another index. So besides these two are the third we're gonna be\ldots no, number of MPI ranks not just each rank, my bad, that's slightly confusing. --- P7 (Student, Technical Advisor)
        \end{quote}

	Intentional aggregation and filtering were common. P2 and P5 alluded to
	aggregation during analysis. P2 is the developer of the statistics
	component which is used to hold derived aggregate statistics.
	P5's role has both developer and analyst elements and thus was
	likely to use the statistics table.

	Reducing dimensionality was implied by P6, who frequently referred to
	simplifying the dataset as part of their workflow:

        \begin{quote}
            you start by looking at the whole thing, right? The entirety of the runtime. And then, you know, you start throwing away things that may be irrelevant like IO\ldots --- P6 (PhD, Technical Lead)
        \end{quote}

	P8 repeatedly indicated that aggregated metrics were most useful to
	them than the granular metrics of the performance data table. P9's
	drawing focused on a particular measurement metric and grouped
	data to reduce those granular metrics to something more
	manageable.

    \theme[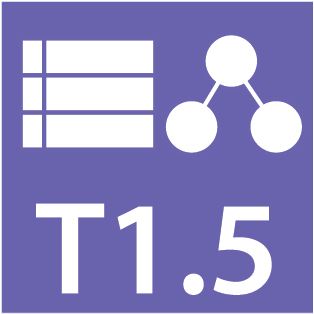]{Participants were divided in their
    conceptions of the heterogeneous dataset as either primarily hierarchical
    or primarily tabular.}
    P1-P3, P7, and P10 generally exhibited a table-centric view,
    to different degrees of explicitness. P1 and P2 never drew the tree
    and referred to ``nodes" only as table indices,
    removed from their tree context. P10 drew the tree but only as an
    amendment to their original drawing after interviewer probing.  P3, P7,
    and P10 all spoke about the tree, but as a supplement providing context to
    the ``real" data of the performance table, with P3 and P7 using relational
    database terminology to describe data operations and relationships.

    Among the participants with a tree-centric view, P4 suggested there are
    many aspects of the data not captured in the reference data model.
    However, they viewed the data as fundamentally hierarchical as ``an
    artifact of how we are defining systems." P5 and P6 used similar language
    that relegated tables as ``storage" or ``transportation layers" between
    the data and useful visualizations, including tree visualizations.  P8
    explicitly described their data as ``a hierarchical grouping of
    elements\ldots" P9 described exploring their dataset as ``\ldots going
    down to specific hierarchy levels," suggesting a hierarchy-first view of
    the data.

\subsection{Data Contexts}
\label{sec:analysts-and-dev-theme}

Participants descriptions of data were embedded in the contexts of their data
sources and analysis workflows.

    \theme[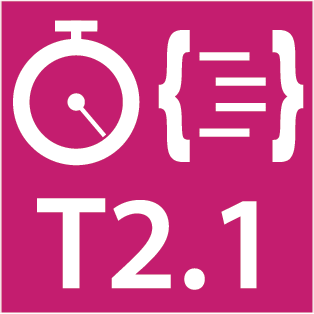]{The participants' mental models relate
    to the sources of data.}
	Participants discussed the source of the data when describing their
	datasets. P4, P5, and P8 reference how measurement
	tools are used to collect data. They speak of decisions made about
	what portions of software to instrument for their research question
	and how to ``tag'' or label them and what metadata to collect. 
	These decisions are subjective and informed by experience,
	but ultimately determine what data comes out and how useful it might
	be for future analysis.

    Data sources were also present in drawings. P2, P4, and P7 used arrows
	to indicate movement between data sources and their model of the
	ensemble dataset. They also discussed verbally the data source as part
	of their model.

    \theme[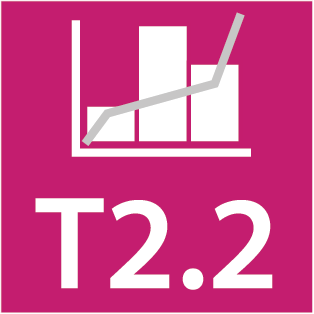]{Participants evoke analysis workflows in
    describing their data.}
	P2, P4-P6, and P8-P10 evoked analysis workflows to support their
	recollection or explanations. P4 and P6 described their data by
	contextualizing it within hypothetical analysis goals. They suggested
	questions which would drive data collection and inform analysis. Both
	P2 and P5 described how data would be derived in a analysis scenario.

	P6 and P8-P10 drew their data as the statistical charts commonly used
	for analysis in the domain. Visible in Fig.~\ref{fig:P6}, 
	these charts are a common plot showing how some
	metric (y-axis) behaves with the change scale of a
	problem (x-axis) or number of resources allocated to it. P6, P8, and
	P10 exclusively used these charts to show their data and found them
	sufficient to describe their data ``in its entirety."

\subsection{Metadata}
\label{sec:metadata-theme}

Participants struggled to acknowledge, communicate, and use metadata despite
its importance to their analysis.

    \theme[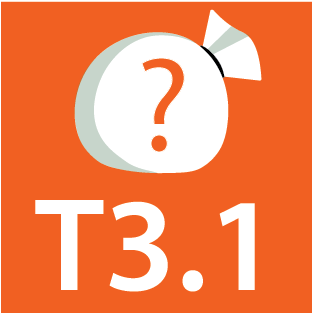]{Metadata is a grab-bag of ``not the performance data.''}
    Participants struggled to define metadata despite uttering numerous
	examples of metadata entries in their heterogeneous dataset. Some
	described it primarily as the context under which measurements were
	collected (P2, P3, P4). Others defined it in a more process-oriented
	way.

	P7 called metadata ``the things that differentiate the runs", implying a
	goal of discrimination. For them, the metadata not only provided
	context about the ensemble's dataset members, but also has use in the
	overall analytical workflow. P5 described the metadata as a mechanism for filtering performance data:

        \begin{quote}
        \ldots we need some metadata in order to make good selection of specific nodes that we are interested in. --- P5 (Student, Analysis Workflow Developer)
        \end{quote}

	The remaining participants did not define metadata explicitly. P6
	echoed P5 and P7's use-oriented casting of metadata by citing specific
	examples of how analysts will use metadata for grouping and
	averaging the data:

        \begin{quote}
        \ldots you're grabbing, you're looking for Machine One and this number of processors, right, to form a little group that you are now averaging and computing you error bars on. Right, and then you grab another one, right? And then you do the same thing for machine two. --- P6 (PhD, Technical Lead)
        \end{quote}

	P8 called metadata the ``basic characteristics of the problem." By
	mentioning a ``problem," P8 suggested that the metadata contains
	information interesting to analysis. 
	
	They further narrowed this definition down to exclude
	certain problem characteristics, carving out specifics of the software
	application's configurations as ``not really metadata." 

	P9 referred to data measurement aspects of metadata, noting ``global
	performance metrics" are metadata. Although these metrics are
	observational measurements collected like the data, they are collected
	at a very low resolution. A typical example is total overall runtime
	for a program. 

	P6 implied metadata while constructing drawings and describing
	analysis goals. They labeling x-axes with ``number of processes" and
	grouped the data by problem sizes.  Metadata was relevant to their
	analysis, but is available and ``just work[ed]" as needed. 

	P8 described metadata as a grab bag, saying:
	\begin{quote}
	\ldots it's very hard to predict what you are going to want to look at\ldots you just kind of throw a bunch of stuff in there. Here's all the things I think might be interesting. --- P8 (Student, Data Analyst)
	\end{quote}

	This description does not imply any strict bounds of the size, type, or representation of data. 

	P10 frequently amended their list of metadata when drawing and
	subsequently answering the task-based questions. When recalling
	a new metadata variable, they added it indiscriminately to a rapidly
	growing list. However, two variables that had special meaning to
	differentiate between member datasets of the ensemble were not placed
	in metadata, instead placed under a separate header, ``attributes''
	which did not correspond to the reference data model.
	
    \theme[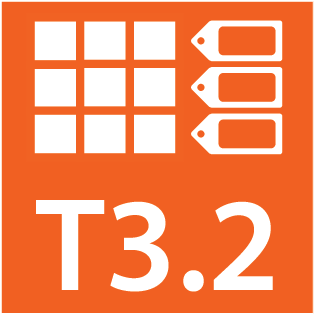]{Participants expressed a desire for more structured metadata.}
	Related to the discussion in T3.1, P5, P8 and P10
	identified deficiencies in how the reference data model
	manages structured metadata. For example, P10 indicated the metadata
	table does not capture the ``attributes'' they considered most
	important to describe experiment conditions, and did not draw these
	attributes with the metadata.

	P8 had more technical issues with the metadata, revealing a similar
	need for more structure and organization:
        \begin{quote}
            there were issues with, with all of the metadata even being there\ldots we had to go through some particularly grueling work to get some metadata columns in there that were constructed from from existing metadata. --- P8 (Student, Analyst)
        \end{quote} 

	This ``grueling work" was further exacerbated by a lack of
	annotations denoting type information about extracted metadata:
	``pulling it out correctly has been a challenge" (P8).

    \theme[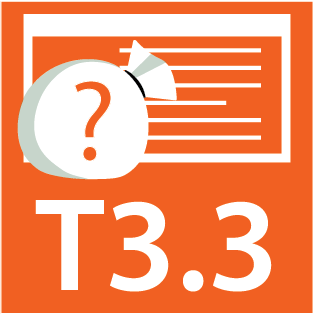]{The value of metadata to participants is highly contextual.}
	Unlike with the performance data (T1.3), participants (P2, P5, P6, P8 and P9)
	had varied sentiments on the value of metadata. In T1.1, several
	participants did not include explicit
	references to metadata in their drawings, suggesting it was not
	important to them. P8 explicitly considered metadata ``useful for
	organizing your data\ldots not useful for analysis." When asked about
	the lack of metadata in their drawing, they said:
        
        \begin{quote}
            And I, you know, I don't think of the metadata really, as I guess, I guess I think of that is a separate piece from the actual timing, tree data, right? --- P8 (Student, Data Analyst)
        \end{quote}

    P5 expressed a similar sentiment of metadata's detachment from data.
	They implied the utility of metadata is limited through their use of a
	modifier, ``only:" 

        \begin{quote}
            \ldots well right now it's kind of, detached in ways, the metadata is only really used to like -- you can look into the metadata, then you can query by it --- P5 (Student, Analysis Workflow Developer)
        \end{quote}
        
	P7 hinted at the use of metadata as optional, saying its relevance
	``depends on what you want."  Less explicitly, P2 and P3 called
	performance data ``the most important piece of [their dataset],"
	relegating metadata to a position of less-than by contrast. 

	Over the course of the interview, mentions of metadata increase as
	participants work through the task-based questions. For example, P1,
	who had the most sparse description and representation, acknowledged
	that something was missing when attempting to answer a task-based
	question, saying ``So that's, that's not in this table." They did
	however, discuss specific values which would have been captured as
	metadata. 

	P8 demonstrated the most significant divergence between their explicit
	opinion of metadata and their discussion of its utility. In the same
	sentence where they dismiss metadata as not useful for analysis, they
	also said ``[Metadata is] useful for things like plots and useful for
	organizing your data." 
	Although plots and organization are not
	explicitly analysis, they are needed to make sense of heterogenous
	datasets. Furthermore, P8 drew a plot to describe their ensemble data.

    \theme[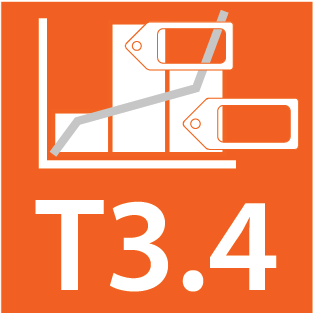]{Metadata supports analysis by facilitating necessary aggregation, filter and grouping operations.}
    \label{sec:filteragg}
	Filtering, grouping, and aggregating data were important to many
	participants. P2 and P4-9 referenced one or more of these
	operations as they walked through their analysis
	narratives.   

	P5, P6, and P8 considered aggregation a pre-requisite to
	understanding the data and performing analysis. P5 implied this by
	describing the data wrangling process to get the mean of a set of mean
	values. P6 implicitly aggregated their data in both of the charts they
	drew to describe the dataset (\autoref{fig:P6}),
	saying ``for this plot, I'm no longer looking at the individual run. So
	I'm looking at the data and the stats frame." P8 expressed that the
	``interesting" data is aggregated saying, ``usually what I'm really
	interested in seeing is something like average time divided by
	something."

	P2, P5-P6, and P8-P9 referenced grouping data during analysis.
	P5 associated grouping with metadata when they describe a hypothetical
	workflow, saying ``if you were to group by, which I do a lot\ldots I'm
	interested in like the number of MPI ranks that are used." MPI ranks
	are a domain example of metadata. P6 discussed grouping data
	frequently. When describing a successful analysis
	session, they said, ``we grouped computation by type\ldots then
	the communication part was\ldots where we were seeing the noise." P9
	and P4 evoked data "slicing" and "cutting" imagery to explain grouping
	and structuring data.
        
	P5 and P6 tied metadata to filtering operations and suggested
	filtering was an important part of their workflows. P5 said, ``I'm only using it to query for stuff." They
	explained the importance of metadata in querying by saying
	``we need some metadata in order to make good selection of specific
	nodes that we are interested in." P6 made explicit reference to
	"filtering" when describing a analysis workflow:

        \begin{quote}
            So when I was describing and now, I'm gonna grab all of the ones for machine one at this scale. That's a filtering operation on that, on the entirety of that data. --- P6 (PhD, Technical Lead) 
        \end{quote}

\section{Discussion}
\label{sec:discussion}

We discuss our research questions, suggest implications for interactive data analysis software, and consider limitations to our study.

\subsection{Research Questions}
\label{sec:rqs}

We present here our findings as they relate to our original research questions.

\subsubsection{RQ1: How well do people remember the aspects of a heterogeneous data structure?}
\textbf{Participants frequently forgot or did not make explicit core parts of
the reference model (T1.1) and dropped key dimensions within the model
(T1.4).} However, with different elicitation prompts, many were able to at
least implicitly recall most parts of the reference data model, which we discuss further
in RQ3. 

In our particular case, we noted that many participants did not mention how
aspects of the model are connected. It is unclear if connections were not
mentioned \note{because they were forgotten, considered apparent without mention, or} something else about their mental models. For
example, P1 said they did not have a model of the data beyond the performance
table which they worked on. For others (P4, P8-P10), connections were not
meaningful in their described mental models. Although connections would be
required to form the data into the shape of their expressed models, those connections were `in the past' of their data narrative and not considered. 

Metadata was another part of the reference data model that was frequently hazy,
implicit, or unmentioned (T3.1). Many participants focused on the performance
data as the ``real data,'' despite needing the metadata for their tasks
(T3.4).

\vspace{1ex}
\subsubsection{RQ2: What factors lead to remembering dimensions/aspects of
the data?}

We \note{identified} several factors \note{that} seemed to influence the recollection of specific
pieces or dimensions in the data.

\textbf{Personal narrative influenced remembering dimensions/aspects of the data.}
\label{sec:recollection2}
    In T2.1 and T2.2 we observed that many participants relied heavily on personal narrative to
    explain their data. Associations between narrative and recollection are
    well-explored in
    cognition~\cite{rubin2003role, schank2014knowledge,simons2022brain}. This
    narrativizing holds for our participants when attempting to recall data.
    P2, P4, and P8 built up a description of the data from the context in
    which it was collected and how it is transformed from ``raw data'' to a
    meaningful reified data model. Others (P6, P10), verbally built the data
    model backwards from its intended outcome in terms of exploratory charts.

    \begin{figure}
        \centering
        \includegraphics[width=.9\columnwidth]{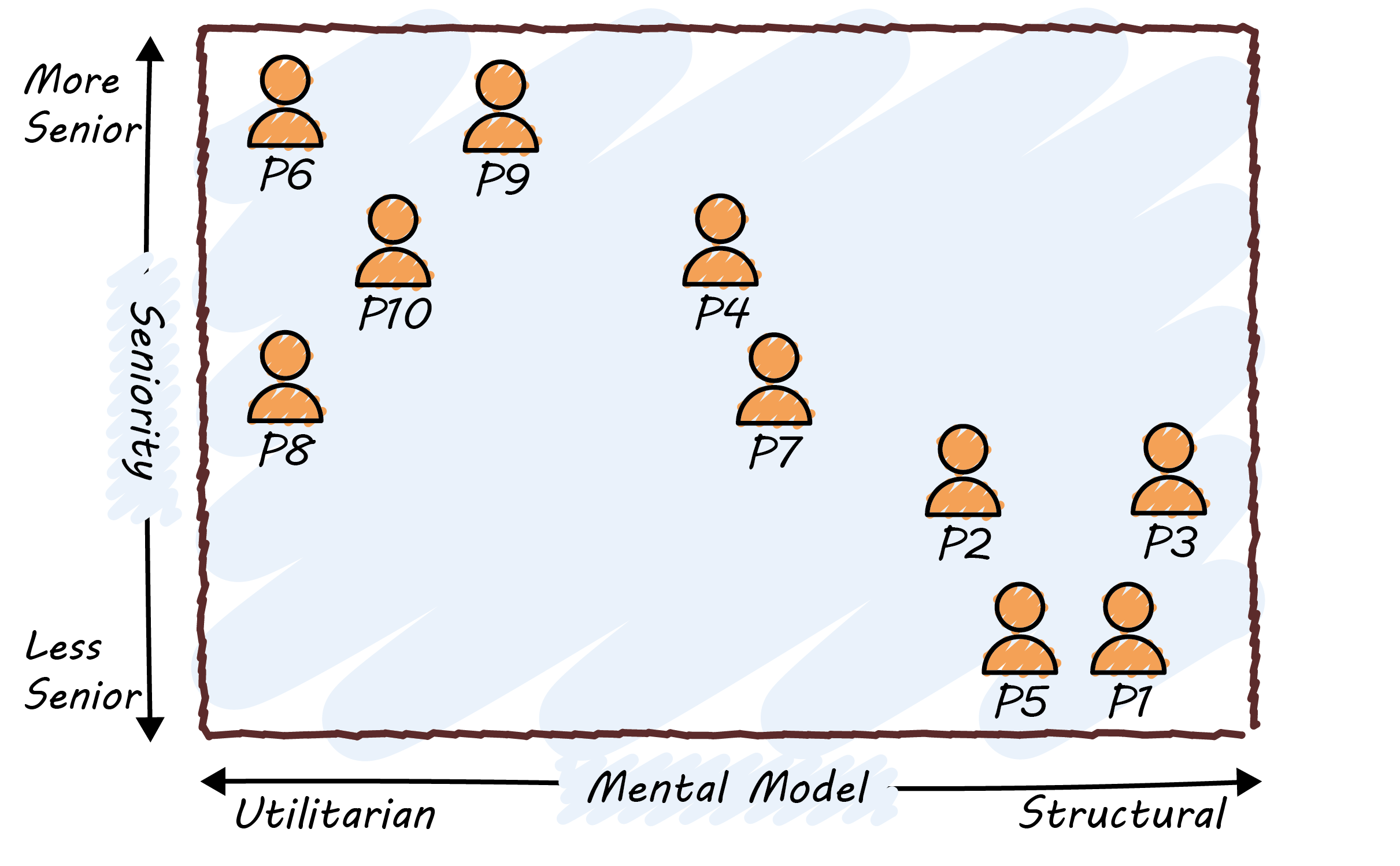}
        \caption{Sketch characterizing participants' mental models related to their seniority. On the x-axis they range from structural (focused on the {\em structure} of the data) to utilitarian (focused on the {\em use} of the data). Participants are placed on the y-axis according their relative seniority---their background and role on the project. More senior individuals held analysis or advisory roles further removed from the day-to-day development.}
        %ALT: A figure showing how participants on two-axis plot characterizing the way they think about their data and how it relates to their seniority. The axes are to the left and bottom of the plot. The x-axis on the bottom is labeled 'Mental Model.' The left side of the x-axis is labeled 'Utilitarian.' The right side is labeled 'Structural.' The y-axis to the right is labeled 'Seniority.' The top of the y-axis is labeled 'More Senior' and the bottom, 'Less Senior.' Icons representing participants are placed on the plot. Each participant is labeled. Less senior participants (P1 through P3, P5) are clustered on lower 'structural' side of the chart and more senior participants (P6, P8 through P10) align with 'Utilitarian' thinking. P4 and P7 are in the middle. 
        \label{fig:spectrum}
    \end{figure}

\textbf{A participant's role on the project influenced what parts of the
model they recalled and/or considered important.} For example, P1 focused on
the table that they primarily develop, despite interfacing with other
developers regarding its use in other parts of {\tt EnsembleAPI}. 

We also noticed there was a spectrum of how participants' mental models in
terms of structural versus utilitarian they were in their descriptions. By
{\em structural}, we refer to those who describe the data more in terms of the
reference data model and its underlying data structure. By {\em utilitarian},
we refer to depictions that focus more on the utility and outcomes of the
analysis the model supports. 

\autoref{fig:spectrum} shows this utilitarian-to-structural spectrum with
regards to the individual participants and their seniority. Participants are placed in this area based on a combination of qualitatively assessed factors, like how much their drawings and verbal responses reference the reference data model. 

Placing our participants on this plot, a few clusters emerge. More senior people on the project exhibit a more utilitarian model and more junior people exhibit a more structural model. There are some exceptions to this and two participants lie roughly in the middle. 

Junior participants were more likely to be development-focused and thus
emphasized the details of the data structure.  Senior participants were more
likely to engage in analysis the tool supports or in interfacing with other analysts, \note{which} may \note{have influenced} their more
outcome-oriented view. This alignment supports the intuition that an
individual's model of data is informed by how they interact with it. It is also
consistent with Sedlmair et al.'s~\cite{sedlmair2012design} pitfall of eliciting tasks from tool developers versus front-line analysts. 

\textbf{Participants recalled more facets of the reference model when
drawing and given explicit tasks.} As discussed in T1.1 and shown in \autoref{fig:timeline},
participants made more parts of their heterogeneous mental model explicit when
drawing or working through tasks over the prior questions that asked them to describe the data and rank importance. 

Existing research in cognition~\cite{dando2013drawing,mattison2015sketching}
suggests a connection between sketching/drawing and memory.  In our context,
this sketch-related recollection may also be attributable to the cognitive
effort required to take something wholly abstract and unintentionally crafted
like a mental model and reifying intentionally it on paper. Drawing may have
also forced previously implicit portions of the model to be made explicit.
Similarly, our tasks were designed to exercise all portions of the model,
leading to recall.

\subsubsection{RQ3: Given a set data structure, how diverse are the mental models of it?}

We observed a diverse set of mental models from our participants. Some
participants prioritized different abstractions in the data structure, like
tables or hierarchies (T1.5). Others expressed their model more in terms of
the output charts than the data (T2.2). As noted in answering our previous two
research questions, the mental models also reflected individual priorities in the parts
of the data structure which were recalled and factors such as role in the project.

Even within clusters of higher level models, there was diversity.  For
example, among the participants who thought in terms of outcome charts, we saw
some that emphasized variance (P6, P9), some which included more dimensions
(e.g., P10 via a multi-line), and others who focused on a simple linear trend,
evoking more the idea of a chart than the character of the data (P8).

This observed diversity, along with participants' concern for data sources and
tasks when expressing their mental models, reflects findings observed by
Williams et al.~\cite{williams2023data}. However, while their study examined small,
unfamiliar datasets \note{where participants had not agreed upon} a "standard" data abstraction, our
study observed this diversity in a larger, more complex case where a reference
data model was known to and had been worked on by the participants for a year.
\textbf{Even with substantial guidance through a known reference model, mental models of the data remained diverse.}

\subsection{Implications}
\label{sec:implications}

The struggles in recalling and describing heterogeneous data we observed along with the
diversity of models suggested by our participants have implications for how we design data exploration interfaces \note{such as data models in scripting APIs}.

\note{\subsubsection{A Tale of Two Hazards: We identify parallel hazards of the mental and reified data models.} 
\label{sec:hazards}
The first hazard we identify highlights a potential failure point in the mental model of a data worker: 

\begin{quote}
    If a data worker does not have a complete and accurate representation of the data in their mind, they may be unable to do the analyses they want, regardless of the reified data model. 
\end{quote}

All mental models risk being hazy, imperfect, and biased representations of the data they represent, however we observed from our interviews that sufficiently inaccurate mental models resulted in errors which would prevent practical analysis of the dataset. Most notably in T1.1, T1.2, T1.3, T1.4, and T3.3 participants evidenced biases towards some parts of the dataset and forgot or eschewed mention of others which they considered less important. When asked to perform concrete analysis tasks, participants noted that a necessary part of the data was not present in their model and had to re-insert it. Furthermore, the need to use the tree or metadata components of their heterogeneous dataset often challenged their explicit rankings of data importance. These instances demonstrate that an inaccurate mental model can prevent analysis even when the reified data model is known and well established.

The second hazard highlights a failure point based on a worker's understanding of a reified data model:

\begin{quote}
    Even if a worker knows what they want to do with data, if they do not understand the reified data model, they may be unable to express their analysis in code.
\end{quote}

In our findings we observed that some participants held concrete conceptions of the types of analyses they wanted to undertake (T2.2, T3.1, T3.4) and expressed their mental model from this perspective. These and other participants were also inclined to simplify or elide parts of the reified data model when discussing or drawing it. They had a clear conception of their desired outcome but did not demonstrate sufficient understanding of the reified data model to explain how to get from a collection of tables attached to a hierarchy to a completed graph.

These two hazards may not in all cases halt analyses but could significantly slow them down. The first hazard can limit people's ability to form the right set of analyses. The second can limit or impede their ability to use a tool or interface.

These hazards suggest that \textbf{the exploratory data analysis workflows of users like our participants can be accelerated by designing data science tools and interfaces that aid in conceptualizing the components of their data both in relation to their analysis and the concrete form of the model encoded into their tools.} 

One way to address the parallel hazards would be through careful design of the reified model to better match the mental models. However, the diversity and incompleteness of models we observed suggests that this approach may be infeasible and potentially limiting. While perfectly matching models may not be possible, there is still opportunity to design tools that both support other data conceptions on top of the reified model as well as design the reified model to better match the conceptions and needs of users. We discuss possible approaches to address these challenges in the rest of this section.

}

\subsubsection{Difficulties in conceptualizing complex and heterogeneous datasets present
a challenge for ad hoc exploratory analysis. Existing and emerging interactive
visualization techniques could be adapted to support these datasets.}  

Although we refer to the model in \autoref{fig:ensemble-api} as a generic
``data model," it is also a representation of the data structure which
organizes {\tt EnsembleAPI}. This means that its fundamental purpose to many
of our participants is a practical one: an object around which they perform ad
hoc analyses with a data science ecosystem of tools. Programmatic analysis
interfaces such as these support exploratory data analysis by giving users
complete flexibility to transform their data. However, as we saw throughout
our findings, particularly T1.1, T1.4, T2.2, and T3.2, our participants'
models did not match the reference data model and they struggled at times to use the more nebulous pieces, such as the dropped dimensions (T1.4), also seen in Williams et al.~\cite{williams2023data} for simpler datasets, or the `grab bag' of metadata (T3.1), hampering their data
investigations. This observation is particularly concerning because if the
team who created the data model and continue to use it forget its components,
it may be even more difficult for its wider intended audience.

To lower the barrier to data analysis in these complex data situations, existing
guidance on visual recommender systems~\cite{crisan2019uncovering,
wongsuphasawat2015voyager, wongsuphasawat2017voyager,
mackinlay2007show,crisan2021gevitrec} may mitigate difficulties. Providing exploratory data visualizations to users may reduce their need to recall the data model. In cases like {\tt EnsembleAPI}, these recommender systems would have to be adapted for the domain specific data model. Doing so would provide support for the
outcome-oriented mental models we observed from the more senior participants in
our study.

Graphical-scripting techniques and
tools~\cite{epperson2023dead,b2,mage,scully2024design} may also help bridge
this difficulty by generating code which describes how visualizations are
produced from the original data model. The combination of code and visual output can bridge the gap between the chart-based mental models we observed and the reified data model by providing a walkthrough
of how one can be made from the other. Additionally, embedded interfaces
that remind the user of the larger data model in a lightweight and
screen efficient way~\cite{epperson2023dead} could be designed to
support multi-component and heterogeneous data models.

\note{
\subsubsection{Structuring data into a model is necessary for developing data exploration interfaces but also imposes technical debt on a project that makes changes to data design difficult. Probing a proposed data model in relation to stakeholder's mental models early may mitigate data model design problems.}
\label{sec:engineering_cost} 

In the project we studied, we observed several cases where the design of the reified data model was in tension with user goals.

For some participants, the data was more granular than they wanted or needed (T1.1, T1.4) or critical components like metadata were not sufficiently structured for ease of retrieval (T3.1, T3.2). Some participants skipped over thinking about the reference data model entirely (T2.2). In terms of meeting the mental models of potential users, this data model evidences some opportunities for improvement. However, the {\tt EnsembleAPI} project has been in development and use for many years now based around this design and thus makes the adoption of changes difficult. The engineering cost required to pivot on a given data design increases as more features are built upon it. An improved initial design could decrease these potential costs.

Recognizing the parallel hazards identified in Section~\ref{sec:hazards} and using them to help guide the initial human-centered data collection required for designing data analysis tools may help produce a more usable reified data model before engineering debt is built up.

By asking stakeholders to describe their data as thoroughly as possible, using a variety of elicitation techniques (sketching, interviewing, and task-based questions), tool designers may be able to better shape a realistic design space for a reified data model. This act of probing users mental models also enables developers and data designers to better understand what data design elements might resonate with their target users inside this reduced design space. 

This preliminary step of inquiry also produces better final products for the real analysts who often end up using data science software and interfaces. According to Liu et al. many "data scientists" are often domain experts primarily doing data science work as needed for their jobs \cite{liu2019understanding}. These "data workers" are often not considered in the design of general analysis software and may be undeserved by the designs of such software's data models and corresponding data-focused interactive interfaces as thier mental data model are unique and needs are different. 

}

\subsubsection{Metadata needs to come into focus.}

    In {\tt EnsembleAPI}, metadata is considered a hazy "grab bag" structure
    that, in many ways reflects the mental models of our participants.
    It is often collected as a basic datatype (string, int, float) and stored
    with only a name identifying its meaning to the user. More complex data
    (maps, arrays) are stored as strings and naively handled like strings in
    analysis software. While {\tt EnsembleAPI}'s support for capturing
    metadata is flexible and easy, it promotes the hazy understanding we
    observed in our participants.

    Metadata is closely related to analysis targets (T3.1) but these targets
    become difficult to reach when methods of accessing the metadata are unclear.
    Furthermore, if such data is not well structured (easily serializable or
    de-serializable) with machine-readable labeling, this complicates the
    process of using metadata even once it's been found. 

    Adding support and encouragement for annotating metadata with semantic
    data types could mitigate problems and provide analysis and visualization
    tools more opportunity to make meaningful recommendations or "smart" default
    displays of the data. For example, ``filepaths" should be displayed in
    different manner  from other strings. Some numerical metadata will be used for
    grouping data and some will be be used for ordering data, as indicated by
    P6's charts in \autoref{fig:sketches}. By understanding what
    characteristics distinguish one of these uses from another, tool developers can make useful metadata nearly as accessible as it
    exists in the user's mind.

    While encouragement on the data collection end can help, interfaces for
    searching, revealing, and annotating the metadata after the fact may help
    in cases where such annotations were not recorded or where new analyses
    suggest viewing metadata in another light. As noted by Bartram et
    al.~\cite{bartram2021untidy}, more human-legible ways to work with
    metadata are needed for analysis.

The haziness of the metadata suggests additional care be taken in
eliciting it. Despite our participants' downplaying of the metadata, it was necessary for the analyses they described.

\subsubsection{A Tale of Many Models: The different models serve different purposes which we can embrace.} 

We were initially concerned when we discovered that P6, P8, and P10 were the most utilitarian-thinking individuals. P6 and P10 were the primary designers of the reference model and yet declined to use it when describing their data, even knowing the interviewers were familiar with it.  P8's analysis tasks also require heavy use of the reference data model, but they also do not describe the data this way. These three individuals aligned in how they think about their data but chose not to use the construct they themselves developed and adopted to describe the data in the abstract. Thus, these individuals had multiple conceptions of the models that they used for different purposes.

Any particular component of the interface, be it an API call, UI element, visualization, or otherwise, may not have to reflect the data as it is stored. Embracing the diversity of mental models may help reveal additional tasks data workers engage in (T2.2) and might indicate that multiple models should be supported by the interface.

\subsection{Study Limitations}

We designed this study around a domain-specific heterogeneous data structure
we were already familiar with so that we could investigate
how people might manage it for analysis. We attempted to recruit all people
familiar enough with it to provide meaningful responses and all but one
volunteered. As such, the number of participants in this study is not large
and we expect there are issues with complex and heterogeneous data beyond the ones we have described as part of this exploratory study.

Interviews were conducted by the first author, a student, with the last
author observing and occasionally adding a follow-up question. This context
may have influenced the way the participants engaged with the interviewer. For
example, the senior participants frequently explain technologies to students they mentor.

We conducted interviews after participants had worked on the {\tt
EnsembleAPI} project for at least a year. All had worked to draft an
academic paper on {\tt EnsembleAPI} in the months before the interview. Our
\autoref{fig:ensemble-api} is based on a figure from that paper. Thus, we expected
them to be familiar with {\tt EnsembleAPI}. However, more time
may be needed to internalize the data model, especially for use rather than
development. Their use of the model may change with time. 

Our observations come from a group of graduate-educated computing
professionals who also act as data workers in performance analysis. The observations focus on reified data model shared across this group. Further studies are necessary to generalize across data models as well as data workers with more diverse backgrounds.

\section{Conclusion}

To better understand and design for data workers using analysis software and heterogeneous datasets, we conducted a qualitative study with ten participants who had been developing and working with the same complex data model, reified through a data science scripting library, for over a year. Our analysis of the resulting data demonstrated  that the participants' mental models were diverse in their structure and frequently diverged from the reified data model they themselves had developed. We identified two parallel hazards associated with mental and reified data models that contribute to data analysis struggles. Based on the themes created during the study analysis, we suggest possible approaches to mitigate these hazards when designing analysis tools and corresponding data models. Awareness of these hazards and early use of human-centered data collection in a development project may help mitigate significant engineering cost associated with changing a data model once established. Additionally, by embracing the diverse mental data models associated with a dataset, we may be able to design interfaces that further help people make sense of complex data and perform the data tasks required to achieve their analysis goals.

%% if specified like this the section will be ommitted in review mode
\begin{acks} 
This work was performed under the auspices of the U.S. Department of Energy by
Lawrence Livermore National Laboratory under contract DE-AC52-07NA27344.
LLNL-CONF-XXXXXX.
\end{acks}

\bibliographystyle{ACM-Reference-Format}
\bibliography{ensdata}

% \appendix % You can use the `hideappendix` class option to skip everything after \appendix

% \section{About Appendices}
% Refer to \cref{sec:appendices_inst} for instructions regarding appendices.

% \section{Troubleshooting}
% \label{appendix:troubleshooting}

% \subsection{ifpdf error}

% If you receive compilation errors along the lines of \texttt{Package ifpdf Error: Name clash, \textbackslash ifpdf is already defined} then please add a new line \verb|\let\ifpdf\relax| right after the \verb|\documentclass[journal]{vgtc}| call.
% Note that your error is due to packages you use that define \verb|\ifpdf| which is obsolete (the result is that \verb|\ifpdf| is defined twice); these packages should be changed to use \verb|ifpdf| package instead.

% \subsection{\texttt{pdfendlink} error}

% Occasionally (for some \LaTeX\ distributions) this hyper-linked bib\TeX\ style may lead to \textbf{compilation errors} (\texttt{pdfendlink ended up in different nesting level ...}) if a reference entry is broken across two pages (due to a bug in \verb|hyperref|).
% In this case, make sure you have the latest version of the \verb|hyperref| package (i.e.\ update your \LaTeX\ installation/packages) or, alternatively, revert back to \verb|\bibliographystyle{abbrv-doi}| (at the expense of removing hyperlinks from the bibliography) and try \verb|\bibliographystyle{abbrv-doi-hyperref}| again after some more editing.

\end{document}